\documentstyle[12pt,moriond,epsfig]{article}
\begin{document}
\heading{LOW RATE STAR FORMATION ACTIVITY DURING QUIESCENT PHASES IN DWARF 
GALAXIES}

\author{F. Legrand $^{1}$, D. Kunth $^{1}$}{$^{1}$ Institut d'Astrophysique 
de Paris, France}{ \ }

\begin{moriondabstract}
Observations of quiescent dwarfs and low surface brightness galaxies
suggest that continuous low rate star formation is likely to occur,
during the quiescent phases between bursts, in dwarf galaxies. We
thus have used a spectrophotometric model to reproduce the observed
abundances in IZw 18, assuming only a low constant star formation
rate. We conclude that such a continuous low star formation rate
cannot be neglected, especially when considering the chemical
evolution of very metal poor objects.
\end{moriondabstract}

\section{Introduction}
Blue compact dwarfs galaxies (BCDG) are still experiencing a
strong star formation event.  Their low metallicity suggests that these
objects are unevolved. The nature and the age of the most underabundant ones
are still controversial: are they ``young'' galaxies forming stars for the
first time or older systems which have evolved very slowly ? Despite of
extensive searches, no local galaxy with a metallicity lower than 1/50
$Z\odot$ has been found, nor massive primordial HI clouds, without optical
counterpart, at low redshift.  These facts could indicate that these objects
are not ``young'' objects experiencing their first episode of star formation.

\section{Abundance profiles in BCDG}
Kunth \& Sargent \cite{KS86}, to explain why no galaxy with a metallicity
lower than 1/50 $Z\odot$ has yet been found, have suggested that during the
starburst phase, metals ejected by the massive stars very quickly enrich the
surrounding HII region. They show that the metallicity would reach a value
up to 1/50 $Z\odot$ in few Myrs. In this scenario, an abundance discontinuity,
between the central starburst region (polluted) and the external regions
(more pristine), is expected. However, recent abundance measurements in IZw
18 (\cite{L98}, \cite{LS98}) and in other starburst galaxies (\cite{KS97} and
references therein) have shown a remarkable homogeneity within the HII
regions over scales larger than 600 pc. On the other hand, these results
appear in contradiction with time-scales required to disperse and mix the
newly synthesized elements, as calculated by Roy \& Kunth
\cite{RK95}. Therefore the most likely possibility is that during a
starburst, heavy elements produced by the massive stars are ejected with
high velocities in a hot phase, and leave the HII region (\cite{DRD97},
\cite{KS97}, \cite{L98}, \cite{TT96}).  This implies that the observed metals
do not come from the present burst, but from a previous star formation event.

\section{Low rate star formation during quiescent phases}
Several studies of IZw 18 have shown that the current burst was not
the first star formation event in the history of that galaxy
(\cite{DH90}, \cite{HT95}, \cite{KMM95}, \cite{GSDS97},
\cite{IT97}). For example, Kunth et al \cite{KMM95} have shown that a
starburst comparable to the present one could be sufficient to account for
the observed abundance. However, such a Star Formation Rate (SFR)
cannot be maintained for a long time without producing excessive
enrichment and consuming all the gas. Starbursts episodes must
therefore be separated by quiescent phases, during which these objects
are supposed to appear as quiescent dwarfs or Low Surface Brightness
Galaxies (LSBG). However, studies of these later objects have revealed
that their SFR was very low but not zero (\cite{vz97}). Thus the
metallicity increases slowly during these quiescent phases. We then
have used a spectrophotometric model to investigate how this low
continuous star formation rate can account for the abundances observed
in IZw 18.

\section{Modeling of IZw 18}

\subsection{Description of the model}

We used the spectrophotometric model described by Devriendt et al 
\cite{DGS98}. The main features of the model are the following:

1) A normalized 1 $\rm M_{\odot}$ galaxy is considered as a one zone closed 
system with instantaneous and complete mixing.

2) The stellar lifetimes are taken into account, {\it i.e.\/,} no
instantaneous recycling approximation is used.

3) The model uses the evolutionary tracks from the Geneva group and the yields
from Maeder \cite{M92}.

4) The stellar output spectra are computed at each age using synthetic stellar 
libraries.

5) We used two typical different IMF (Salpeter and Scalo) described as a 
power law $\phi(m)=a.m^{-x}$ in the mass range 0.1-120 $\rm M_{\odot}$.

6) The evolution of several chemical elements (C,O,Fe), the total
metallicity, the mass of gas, is followed in detail. 

We adopted a value of $\rm 10^{8} \ M_{\odot}$ for the initial mass of
gas in IZw 18.  A constant SFR was adjusted to reproduce the observed
oxygen abundance in IZw 18 after 16 Gyrs.  Finally, a strong star
formation episode at 16 Gyrs (with a SFR of $0.04 \ M_{\odot}/yr$ over 50
Myrs) was added to reproduce the current burst and compare
the predicted colors with the observed ones.

\subsection{Results}

\begin{figure}
\psfig{figure=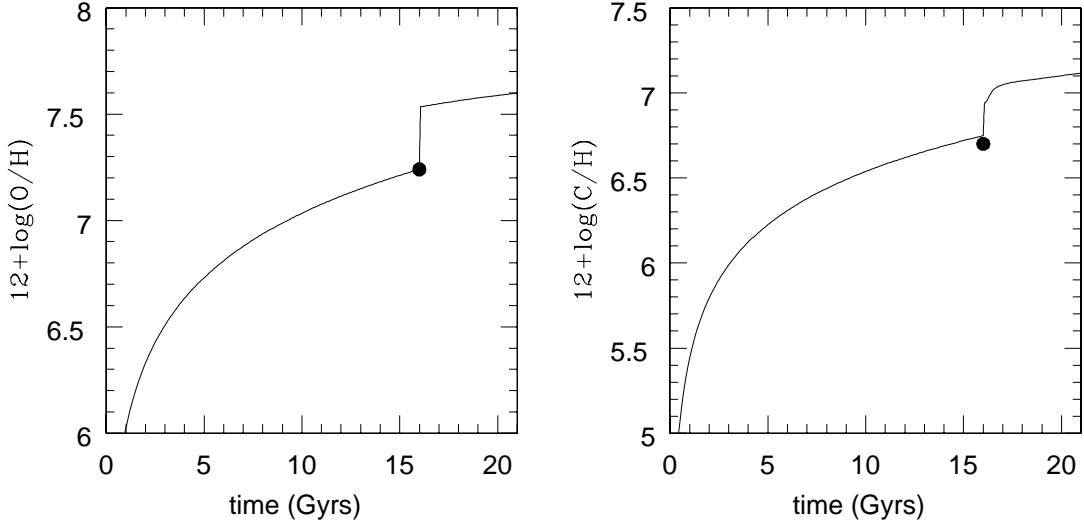,height=7.cm,clip=,angle=0}
\caption[]{Evolution of abundances with time assuming a Scalo IMF. 
The measured abundances in IZw 18 are represented by a 
dot at 16 Gyrs.} \label{fig:abundances}
\end{figure}

We found the required continuous SFR to be around $ \rm 10^{-4} \
M_{\odot}.yr^{-1}$ in order to reproduce the observed oxygen abundance
in IZw 18.  This SFR is ten times lower than what is typically
observed in LSBG and 400 times lower than the present one.  The carbon
abundance \cite{DH90} observed in IZw 18 is well reproduced as shown in 
fig \ref{fig:abundances}.

\begin{figure}
\psfig{figure=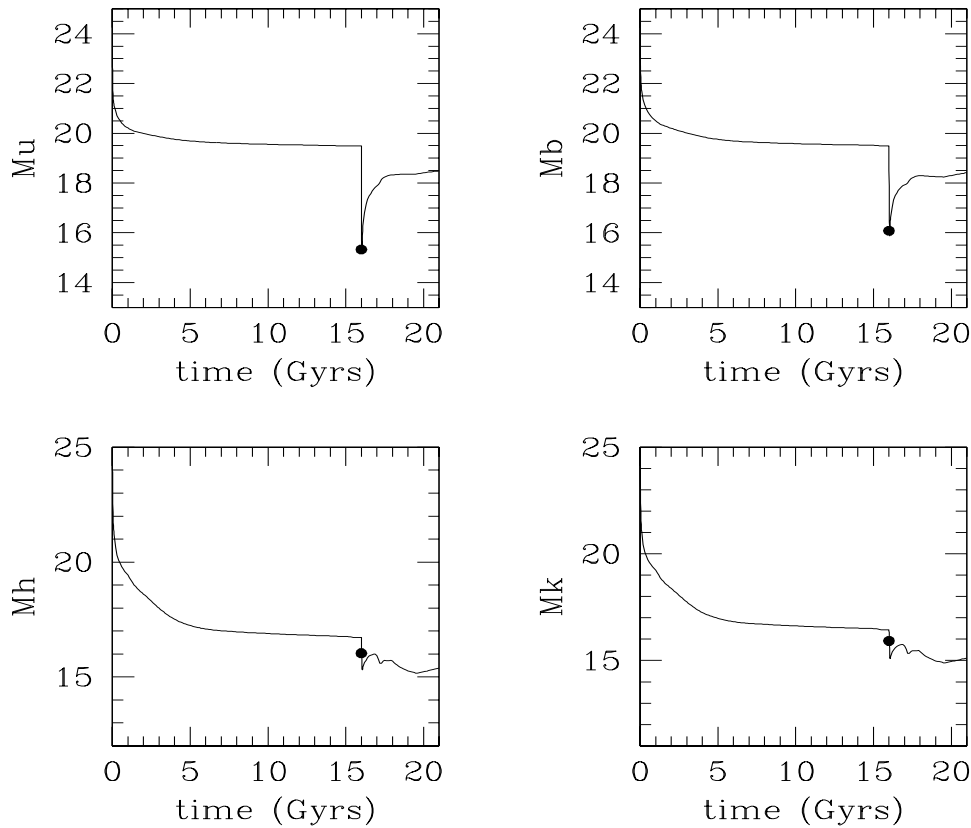,height=11cm,width=16cm,clip=,angle=0}
\caption[]{Evolution of colors with time assuming a Scalo IMF. 
The measured colors (\cite{T83} \cite{H77}) in IZw 18 are represented 
by a dot at 16 Gyrs.} \label{fig:colors}
\end{figure}

The resulting  magnitudes shown in fig \ref{fig:colors}, assuming a 
distance of 10 Mpc for IZw 18, are compatible with
the measurements of Thuan \cite{T83} and Huchra \cite{H77}.

Van Zee et al \cite{vz98} reported a size for the HI envelope (at a column
density of $10^{20}$ atom $\rm cm^{-2}$) in IZw 18 of 60''x45'' {\it i.e.\/,}
3x2.3 kpc. Using this value and assuming that the underlying stellar
component due to the constant star formation process is uniform, we have
evaluated the surface brightness produced by this old population to be 29.3
$\rm mag.arcsec^{-2}$ in the B band and 26.3 $\rm mag.arcsec^{-2}$ in
K. Further observations should be performed in order to observe this faint
population in region far from the central burst.

\section{Conclusions}

This modeling indicates that IZw 18 could undergo its first ``burst'' of star
formation, but would not be a ``young'' galaxy in the sense that a mild
process of star formation already started a long time ago. If these kind of
objects are not ``young'', this would reconciliate with the negative
detections of massive primordial HI clouds, without optical counterpart, at
low redshift, required for their late formation, explaining why no galaxies
with a metallicity lower than 1/50 $Z\odot$ have been found.

We have shown that, low rate star formation is likely to occur during the
quiescent phases between bursts in dwarfs galaxies and cannot be neglected,
especially when dealing with the chemical evolution of very metal poor
objects. A more detailed version of this work will be presented
elsewhere (\cite{LS98}).


%
\vfill

\begin{moriondbib}

\bibitem{DRD97} Devost D., Roy J-R., Drissen L., 1997, \apj {482} {765}
\bibitem{DGS98} Devriendt J., Guiderdoni B., Sadat R., 1998, {\it in preparation}
\bibitem{DH90}  Dufour R.J., Hester , 1990, \apj {350} {149}
\bibitem{GSDS97}  Garnett D.R. et al, 1997, \apj {481} {174}
\bibitem{H77} Huchra J.P, 1977, \apjs {35} {171}
\bibitem{HT95} Hunter D.A., Thronson H.A., 1995, \apj {452} {238}
\bibitem{IT97} Izotov Y. \& Thuan T.X., 1997, \apj {}{in press}
\bibitem{KS97} Kobulnicky H.A. \& Skillman E.D. 1997, \apj {489} {636}
\bibitem{KS86} Kunth D. \& Sargent W.L.W., 1986, \apj {300} {496}
\bibitem{KMM95}  Kunth D., Matteucci F., Marconi G., 1995,  \aa {297} {634}
\bibitem{L98} Legrand F., 1998, {\it Proceedings of workshop at U. Laval,
Quebec, ASP conf. series}
\bibitem{LS98} Legrand et al, 1998, {\it in preparation}
\bibitem{M92} Maeder A., 1992, \aa {264} {105}
\bibitem{RK95} Roy J.R. \& Kunth D., 1995, \aa {294} {432}
\bibitem{TT96} Tenorio-Tagle G. 1996, \aj {111} {1641}
\bibitem{T83} Thuan T.X., 1983, \apj {268} {667}
\bibitem{vz97} van Zee L. et al, 1997, \aj {113} {1618}
\bibitem{vz98} van Zee L., Westpfahl D., Haynes M.P., 1998, \aj {115}{1000}

\end{moriondbib}
\end{document}